\documentclass[preprint,floatfix]{revtex4}
\usepackage{amssymb}
\usepackage{amsmath}
\usepackage[usenames]{color}
\usepackage{mathrsfs}
\usepackage{euscript}
\usepackage{graphicx}
\usepackage{graphics}
\usepackage{bm}
\usepackage{color}
\usepackage{pdfpages}

\begin{document}

{\Large\bfseries\mathversion{bold}
~~~~~
 \par}%
\vspace{8mm}


# Multiple orbital contributions to molecular high-harmonic generation in an asymmetric top

Limor S. Spector[1,2*], Shungo Miyabe[2,3], Alvaro Magana[4], Simon Petretti[4], Piero Decleva[5], Todd Martinez[2,3], Alejandro Saenz[4], Markus Guehr[2], and Philip H. Bucksbaum[1,2]

[1]Applied Physics Department, Stanford University, Stanford, CA 94305, USA
[2]Stanford PULSE Institute, SLAC National Accelerator Laboratory, Menlo Park, CA 94025, USA
[3]Chemistry Department, Stanford University, Stanford, CA 94305, USA
[4]Institut für Physik, Humboldt-Universität zu Berlin, Newtonstr. 15, 12489 Berlin, Germany
[5]Dipartimento di Scienze Chimiche e Farmaceutiche and CNR-IOM Democritos and INSTM, Università di Trieste, Via L. Giorgieri 1, I-34127 Trieste, Italy

**High-order harmonic generation (HHG) in aligned linear molecules can offer valuable information about strong-field interactions in lower-lying molecular orbitals, but extracting this information is difficult for three-dimensional molecular geometries. Our measurements of the asymmetric top $SO_2$ show large axis dependencies, which change with harmonic order. The analysis shows that these spectral features must be due to field ionization and recombination from multiple orbitals during HHG. We expect that HHG can probe orbital dependencies using this approach for a broad class of asymmetric-top molecules.**

High-order harmonic generation (HHG) has shown promise as a sensitive probe of molecular orbitals based on a dependence of the strong-field interaction with electrons for atoms or linear molecules[1-3]. In molecules, this is typically accomplished by triggering a HHG wave packet along the axis of an aligned molecule. Ionization or recombination from multiple valence orbitals in these linear molecules can be interrogated by observing orbital differences in the one rotational dimension[4-6]. In this Letter, we show that multiple orbital contributions to high harmonic generation can be discerned even for three-dimensional molecules like asymmetric tops. This is done without aligning the sample in three dimensions or using a retrieval algorithm[1]. Contributions from different axes are decoupled by making use of the full time-dependant revival pattern in a manner conceptually related to the well-established technique of rotational coherence spectroscopy[7,8]. The data show large axis dependencies, which change with harmonic order. We use molecular symmetries to determine that these dependencies arise from a wave packet emanating largely from the highest occupied molecular orbital (HOMO) at low harmonics and lower-lying valence states at higher harmonics.

HHG studies of electron orbitals in molecules rely on laser-induced molecular alignment. This is critical to decoding a molecular HHG spectrum since an isotropic molecular ensemble cannot reveal directional information. For a linear molecule, periodic transient alignment is straightforward through impulsive transient stimulated rotational Raman scattering with a linearly polarized laser pulse[9-14].

Asymmetric-top molecules have three different moments of inertia and a complicated polarizability tensor. For such molecules an impulse delivered to an asymmetric top molecule excites free rotation about multiple axes, with up to five different revival constants[7]. There have been several attempts to construct multiple-pulse or pulse-shaping protocols to align one or more axes of an asymmetric top molecule in the laboratory, but three-dimensional alignment has remained challenging[15-19]. Moreover, the rotational coherences of an impulsively-driven

asymmetric top are markedly different from the well-studied rotational revivals of simpler linear molecules. There is no time delay following an impulse for which there is a cleanly separated alignment of any principal axis in an asymmetric-top molecule.

Here we instead employ a new alignment-decomposition approach to extract axis-based alignment information from asymmetric-top molecules in the molecular frame[8]. We decompose the full time-delayed rotational quasi-revival trace from each harmonic into amplitudes and phases corresponding to the quasi-revival pattern for each axis in the molecule. As shown in Fig. 1, we combine this approach with an understanding of the molecular symmetries to decode the harmonic spectrum in $SO_2$. An extremely important experimental conclusion in the diatomic molecules $CO_2$ and $O_2$ has been that the presence of a node in the electron wave function along an axis leads to a suppression of HHG along that axis[6,20]. For $O_2$ and $CO_2$ there is such a node along the internuclear axis, which is easy to align, and so the suppression effect on HHG along the axis of the molecule has been studied extensively. A similar and compelling dependence of HHG on orbital shape should exist in our present experiments on an asymmetric-top molecule.

The $SO_2$ molecule has $C_{2v}$ symmetry, and therefore can support only four different types of molecular orbitals, which correspond to the four irreducible representations of the group. This is in contrast to linear molecules such as $CO_2$, which have an infinite number of irreducible representations each with its own pattern of nodal surfaces.

The irreducible representations of the $C_{2v}$ symmetry group are named $a_1$, $a_2$, $b_1$, and $b_2$. They behave differently with respect to the rotation and inversion symmetries, which necessitate the presence of nodal planes containing different principal axis. The $a_1$ orbitals are symmetric with respect to reflection about all three axes, and therefore should have no nodal suppression along any principal axis. The $a_2$ orbitals are antisymmetric about the a-b plane and about the b-c plane, so that there are nodes along all three axes. The $b_1$ orbitals have a b-c nodal plane, and the $b_2$ orbitals have an a-b nodal plane. Therefore, if we can determine along which axis there is HHG signal enhancement or suppression, we can determine which types of orbitals contribute to the signal. For example, HHG enhancement along the b axis must be due to an $a_1$ state, but enhancement along the a axis while there is suppression along the b axis indicates a $b_1$ or an $a_1$ orbital is responsible. $SO_2$ is thus an ideal test asymmetric-top molecule to look for multiple orbital contributions to an HHG signal.

Our experimental scheme uses an impulsive alignment laser pulse that excites Raman coherences, but has insufficient intensity to ionize a cooled ensemble of $SO_2$ molecules (pulse duration 130 fs pulse, central wavelength at 800 nm, repetition rate 1 kHz, intensity in the $10^{13}$ W/cm$^2$ range). We also generate harmonics using a commercial Ti:sapphire laser (pulse duration 30 fs pulse, central wavelength at 800 nm, repetition rate 1 kHz). Both beams are linearly polarized parallel to one another. The harmonic beam is focused to an intensity of approximately 2.6 x $10^{14}$ W/cm$^2$ and probes the sample at various time delays between 0 and 30 ps. Both beams are focused below the $SO_2$ gas jet. The molecular sample is sufficiently above the laser focus to allow molecules to cool to 240 ± 20° K. We calculated the temperature by assuming a supersonic expansion and using the equation of state for an ideal gas with backing pressure of 1 bar and a heat capacity ratio of 1.29 for $SO_2$[21]. The $SO_2$ is free to rotate; this results in transient alignment revivals of the HHG signal, which is sensitive to the electronic orbital structure along the aligned axis. Harmonics between 20 and 70 eV pass through an aluminum filter (thickness 100 nm) onto a flat-field diffraction grating spectrometer. The signal is amplified with a microchannel plate and detected on a phosphor screen.

Figure 2 shows a full experimental spectrum of the rotational coherence transients of $SO_2$ for harmonics 19-33. At the full J-type and C-type rotational quasi-revivals of the molecule at 26 ps and 28 ps and at the half J-type quasi-revival at 13 ps, harmonics 19 and 25 display a harmonic modulation with the opposite sign in Fig. 2. At times when harmonic 19 is enhanced, harmonic 25 is suppressed. The full harmonic spectrum shows that this reversal in HHG efficiency happens gradually, over several harmonics.

This effect is markedly different from previous investigations of rotational revivals in diatomic molecules. There is no time delay corresponding to any alignment axis in $SO_2$. For example, at the prominent J-type revival at 26 ps, the alignment cosines of the three principal axes are 0.37, 0.32 and 0.31. This most prominent revival therefore cannot be used to determine whether the HHG emanates from any one of these directions in the molecule, or from all three, nor can this revival demonstrate the relative phases of HHG production from these different axes. The full coherence pattern, on the other hand, reveals all of this information.

We quantify this reversal in HHG efficiency by employing the alignment decomposition technique. The data from each harmonic is fit to a combination of three single-axis alignment patterns, which represent alignment as seen from each molecular axis[8,18]. This allows us to obtain coefficients representing the contribution to harmonic emission from each molecular axis independently.

Results from the alignment-decomposition analysis are shown in Fig. 3. Electric field amplitudes in Fig. 3a show that the 19[th] harmonic is dominated largely by contributions from the b axis, but later harmonics show a stronger contribution from the a axis of the molecule. The data further indicate that the transition in the enhancement occurs over the 21[st] to 23[rd] harmonics. Fig. 3b shows the phases we obtain via the analysis; only the relative phase of the three coefficients is meaningful because of the absence of external phase information. Combined with the magnitude information these help explain the HHG intensity seen in the experiment. For example, while in the 19[th] harmonic the phases cause the amplitudes to add constructively, in the 23[rd] harmonic the dampened harmonic signal that results is due to three strong but destructively interfering channels (see Fig. 3c).

Thus a question remains: why does high harmonic emission switch from arising predominantly from the molecular b axis at the 19[th] harmonic to arising predominantly from the molecular a axis at the 25[th] harmonic and above? To explore this, we perform calculations in $SO_2$ modeling the ionization and recombination steps of HHG, which are the times in the semi-classical model of HHG[22-24] when the electron interacts with the molecular orbitals.

We briefly outline our model for high-harmonic recombination; for details see Refs.[25-27]. We compute fixed-nuclei photoionization cross-sections by using the complex Kohn variational method[25]. For final ionic target states $\chi_\Gamma$ labeled by $\Gamma$ and channel functions $F$ describing the photoionized electron, we can write the final-state electronic wavefunction as

$$\Psi_{\Gamma_0 l_0 m_0} = \sum_{\Gamma l m} A(\chi_\Gamma F_{\Gamma l m \Gamma_0 l_0 m_0}) + \sum_i d_i^{\Gamma_0 l_0 m_0} \Theta_i \quad (1)$$

where $A$ is the antisymmetrization operator and the $\Theta_i$'s are $N$ electron correlation terms. This wavefunction describes the production of photoions in a specific cation state $\Gamma_0$ with final electronic angular momentum $l_0 m_0$. We then construct the photoionization cross sections in the molecular frame by using the matrix elements

$$I^\mu_{\Gamma_0 l_0 m_0} = <\Psi_0|r^\mu|\Psi_{\Gamma_0 l_0 m_0}>,$$

where $r^\mu$ is a component of the dipole operator, which we evaluate here in the length form and $\Psi_0$ is the initial-state wave function of the neutral $N$-electron molecule.

To model strong-field ionization, we describe the molecule within the density-functional theory (DFT) by adopting the LB94 exchange-correlation potential that assures the proper asymptotic behavior for Coulomb systems. We solve the time-independent Kohn-Sham equations using special basis functions. This basis comprises spherical basis functions in which the radial part is expanded in B splines and the angular part is expanded in spherical harmonics. The spherical basis functions are locally defined within non-overlapping spheres around each atom and one big central sphere that overlaps with all other spheres. Applying fixed-boundary conditions and subsequently diagonalizing the resulting, field-free Hamiltonian matrix yields a set of bound states as well as box-discretized continuum states that we use as a basis-set expansion for solving the time-dependent Schrödinger equation (TDSE) of the molecule interacting with a laser pulse[28]. In contrast to the standard frozen-electron approximation in which all but one electron is kept in an unchanged field-free orbital, here we adopt a variant of the single-active-electron approximation in which all orbitals are propagated, but within the single-determinant approximation. This leads to gauge invariance and fulfillment of the Pauli principle at any instant of time[29-31]. We obtain the ion yield after the laser pulse by summing the population of all field-free continuum states (or by $1 - P_{bound}$ where $P_{bound}$ is the population of all bound states). Since the photoionization cross-sections and the strong-field ionization yields are calculated with very different approaches with respect to both the underlying electronic-structure model and the adopted type of basis functions, the photoionization cross-sections are additionally calculated within the approach used in the strong-field calculation. Inverting the field-free Hamiltonian matrix $H$ (specifically, the matrix $A^+A$ with $A=H-ES$ and the overlap matrix $S$) within the so-called free-boundary approach, in which the value at the boundary of the central radial basis is left variable while the energy $E$ is fixed yields a proper molecular continuum final state. The photoionization cross-sections can now be extracted from the dipole transition-matrix elements between the given initial state (described by one of the occupied orbitals) and these final continuum states[32]. The agreement between the cross-sections obtained by both methods is generally very good and the conclusions below do not change if using either of them (see Supplemental Material for details)[33].

We weight the photoionization cross-sections, which are proportional to the recombination cross-sections, from the complex Kohn calculation with the strong-field ionization yields obtained from the full TDSE calculation to obtain a rough model of the link between HHG and the molecular orbitals in $SO_2$. Calculations of the photoionization cross-section without a weighting with the field ionization yields do not match the data (see Supplemental Material for more details). The weighted results are shown in Fig. 4, where the theory (on the bottom) is compared to the results of the experiment (on the top). The first column shows the valence orbitals of $SO_2$ (HOMO through HOMO-5) with the orbital ordering obtained from Ref. [34].

Information in Fig. 4 is conveyed through vector models drawn to scale corresponding to harmonic emission from each of the three molecular axes. Thus, the experimental row shows, as before, that the b-axis emission is dominant for harmonic 19, while the a-axis emission is dominant for harmonic 25. The combined theory calculations are separated by orbital and by energy; we can therefore use the calculations to identify which orbitals have an axis-emission pattern that most closely matches the experimental pattern for a particular harmonic.

Calculations for the $8a_1$ HOMO orbital show a strong contribution from the b axis to both harmonic energies. Yet calculations for several of the lower-lying orbitals, including those for the $5b_2$, $4b_2$, $7a_1$ and $2b_1$ orbitals, show that the b-axis signal is suppressed compared to the a-axis one for both the 19th and 25th harmonics. A similar comparison for all harmonics and orbital energies shows they follow a similar pattern. We therefore find that the 19th harmonic sees a dominant contribution from the HOMO, in agreement with previous studies[4]. Higher harmonics, however, retain a larger contribution from the $5b_2$, $4b_2$, and $7a_1$ orbitals. While the relative importance of the different orbitals and thus also the different directions depends on intensity[31], the main conclusions are valid over a rather large intensity window (see Supplemental Material for details)[33].

The contributions from multiple orbitals to the HHG signal in $SO_2$ result from the orbital electron density. Dominant orbitals are generally expected to have more of their electron density in the direction of the laser field[31,35-39]. Here, the HOMO has a big lobe that gives contributions in the b-axis direction, which are vastly diminished in the lower-lying orbitals. Conversely, some of the lower-lying orbitals have a large amount of electron density in the a-axis direction that is not present in the HOMO. This reasoning also explains why the $8a_1$, $5b_2$, $4b_2$, and $7a_1$ orbitals contribute much more strongly to the overall ionization and recombination signal than the $1a_2$ and $2b_1$ orbitals. The most polarizable directions of $SO_2$ are the b- and a- axis directions. The orbitals that contribute most have more electron density in these two directions.

Note that we could have arrived at a similar conclusion by considering molecular symmetry instead of the theory. Nodal planes due to symmetries imply that a dominant signal from the b axis as in the case of the 19th harmonic is due to an $a_1$ orbital like the HOMO. Alternatively, enhancement along the a axis and suppression of signal along the b axis means that either an $a_1$ or a $b_2$ orbital contributes, like the $5b_2$, $4b_2$, and $7a_1$ orbitals that we found contributed. The $a_1$ orbitals may emit along any direction, but the $8a_1$ HOMO orbital has a very large lobe of electron density along the b direction, while the $7a_1$ orbital has a very small lobe in that direction.

Our observation of a marked spectral inversion signal arising from ionization of multiple orbitals is striking in several ways. Here we see the effects of high-harmonic interference from multiple orbitals in asymmetric-top molecules for the first time. This shows that use of HHG to probe electron orbitals and obtain meaningful information is not limited to linear molecules or molecules with a one-dimensional signal. Moreover, it has been reported previously several times in literature that the HHG spectrum is determined largely by the photoionization cross-section[35,40,41], but we show here that such a calculation is sometimes insufficient.

An implication of our results is the advancement of orbital probing using HHG into more complex molecules. We address here the phase-matching limitation HHG has of interference from multiple distinct rotations in asymmetric top molecules. However, HHG spectroscopy in increasingly complex molecules is limited not only by the inherent interference problems caused by having a large aggregation of molecules rotating at different speeds in different directions, but also by the necessity to ionize the molecules sufficiently to produce high harmonics in the first place. As molecules become larger, it becomes more difficult to ionize them and the high harmonic cutoff is lowered, leading to fewer harmonics produced. It is in part due to this limitation that HHG has been conducted only on a handful of very small molecules. Our work here shows a structural inversion at much lower energies than have been seen in previous studies[4,6]. This is evidence of a countervailing effect, which is that as molecules become more complex, their structures become grosser and these larger sizes are visible to lower energy

harmonics, which have a longer wavelength. Thus, a corollary of our result is that although for larger molecules higher harmonics cannot necessarily be produced, they are not required, because large symmetry differences may exist between valence and subvalence orbitals at lower harmonic energies. A second corollary is that more complex molecules generally possess less symmetry and thus fewer types of orbitals, which makes the molecules easier to analyze.

This result points the way to further experiments: using the alignment decomposition technique allows us to extract both the magnitude and phase experimentally along all three dimensions and establishes a set of basis vectors. To obtain a fully three-dimensional image of the molecule as seen by each harmonic, linear combinations of the appropriate basis vectors corresponding to each angle need to be taken such that the full orbital shape can be inferred.

FIGURE 1: The alignment decomposition method. Interference from the three molecular axes normally makes it difficult to disentangle the high-harmonic signal seen in the spectrum, but this method allows us to obtain the harmonic emission in the laboratory frame. We extract axis-dependent information from all axes using the time-evolution of the HHG following a single impulse. This allows us to obtain information pertaining to the molecular orbitals. $SO_2$ has four types of orbitals: $a_1$, $a_2$, $b_1$ and $b_2$. $a_1$ orbitals have no nodes along the molecular axes, $a_2$ orbitals have two nodal planes along all molecular axes, and b-type orbitals have one nodal plane each. Since HHG is stronger for a larger dipole and weaker along a node, we can use the signal extracted along the principal molecular axes to determine which orbitals are dominant for a particular harmonic.

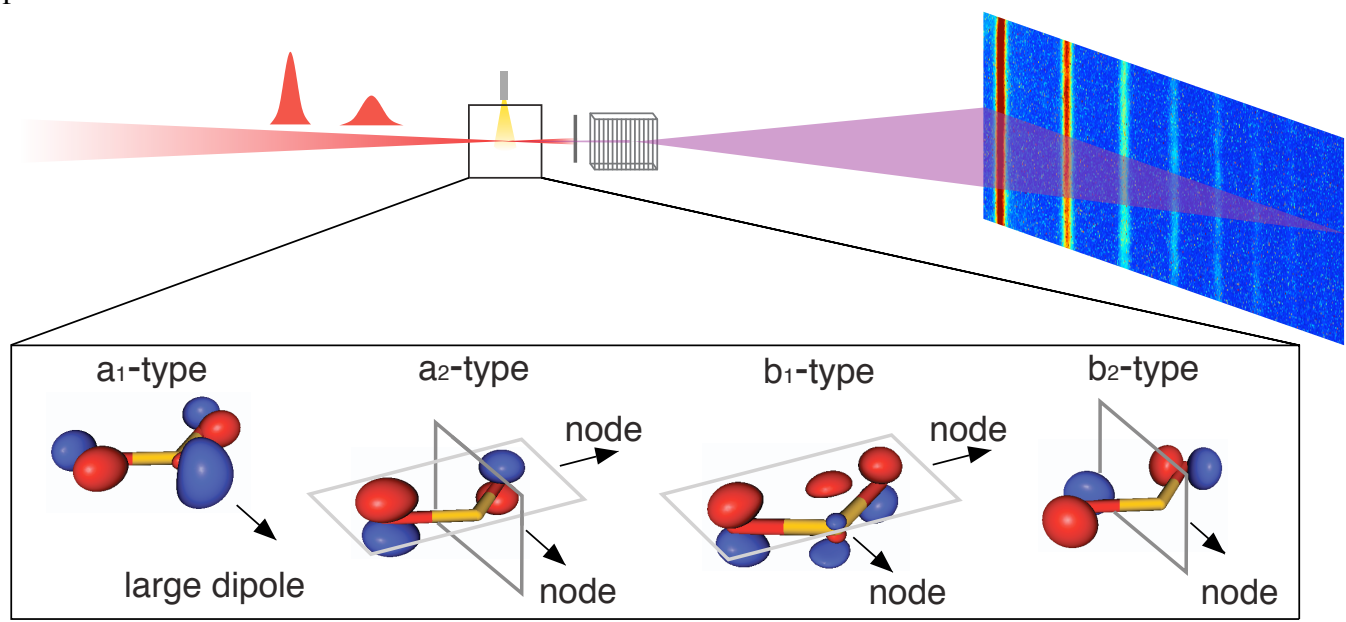

FIGURE 2: Data from the 19th-33rd harmonic in SO$_2$ over the first 30 ps. At the full J-type and C-type rotational revivals of the molecule at 26 ps and 28 ps and at the half J-type rotational revival of the molecule at 13 ps, harmonic 19 and 27 display a harmonic modulation with the opposite sign in Fig. 2. At times when harmonic 19 is enhanced, harmonic 27 is suppressed. The full harmonic spectrum shows that this reversal in HHG efficiency happens gradually, over several harmonics. The modulation of the sign happens gradually, with a dampening of harmonic emission at the 23$^{rd}$ harmonic.

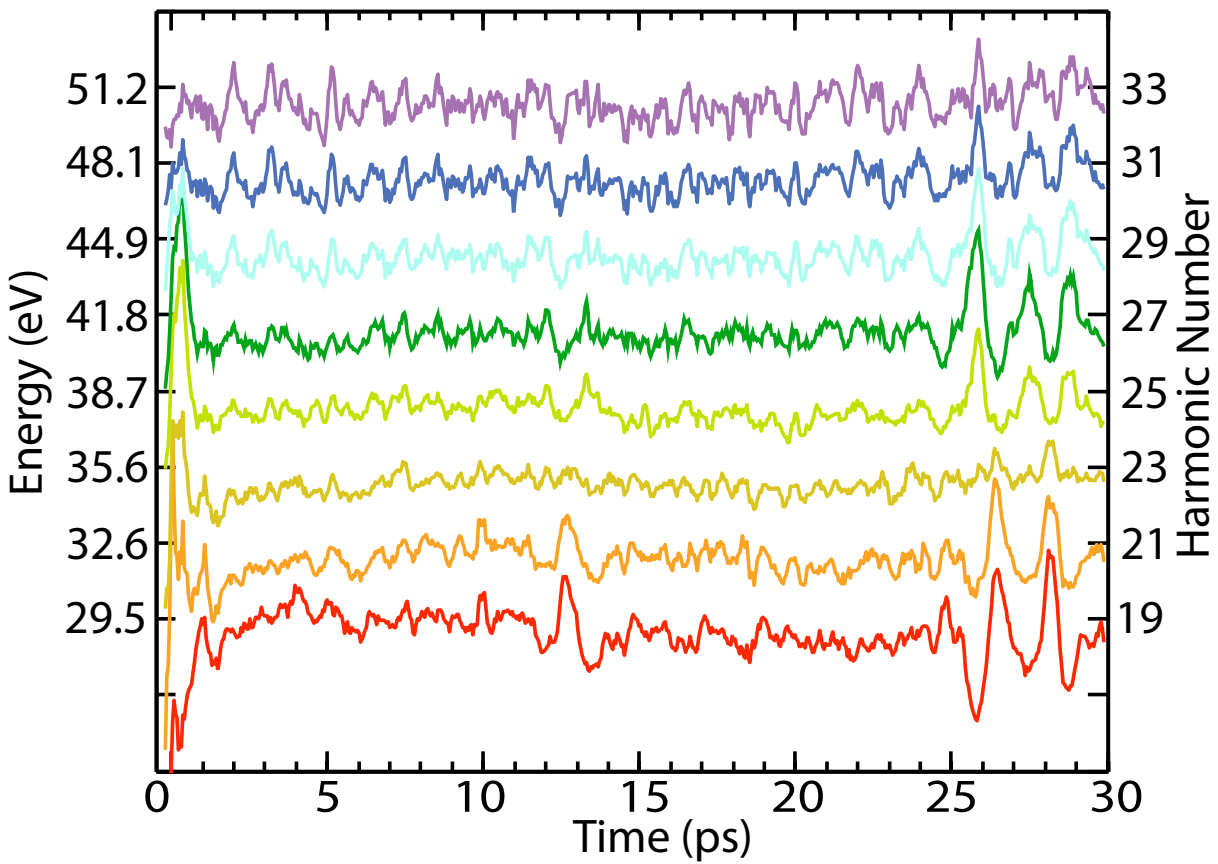

FIGURE 3: Results of the alignment decomposition fit. (a-b) Amplitude and phase of the electric field along all three molecular axes as obtained from analyzing the data. The amplitudes show that the transition in the enhancement occurs over the 21$^{st}$ to 23$^{rd}$ harmonics. The 19$^{th}$ harmonic is dominated largely by contributions from the b axis, but later harmonics show a stronger contribution from the a axis of the molecule. (c) Phasor plots to visualize the magnitude and phase for harmonics 19 and 23. In harmonic 19, the phases are similar and the harmonic emission from each of the three molecular directions adds constructively. In contrast, the dampening of the signal along the 23$^{rd}$ harmonic can be understand as interference of harmonic emission from each of the three molecular directions. (d) Model of the SO$_2$ molecule, displaying the molecular axes. The a axis connects the oxygen atoms, the b-axis is the symmetry axis of the molecule (C$_{2v}$ axis) and the c axis is perpendicular to the molecular plane.

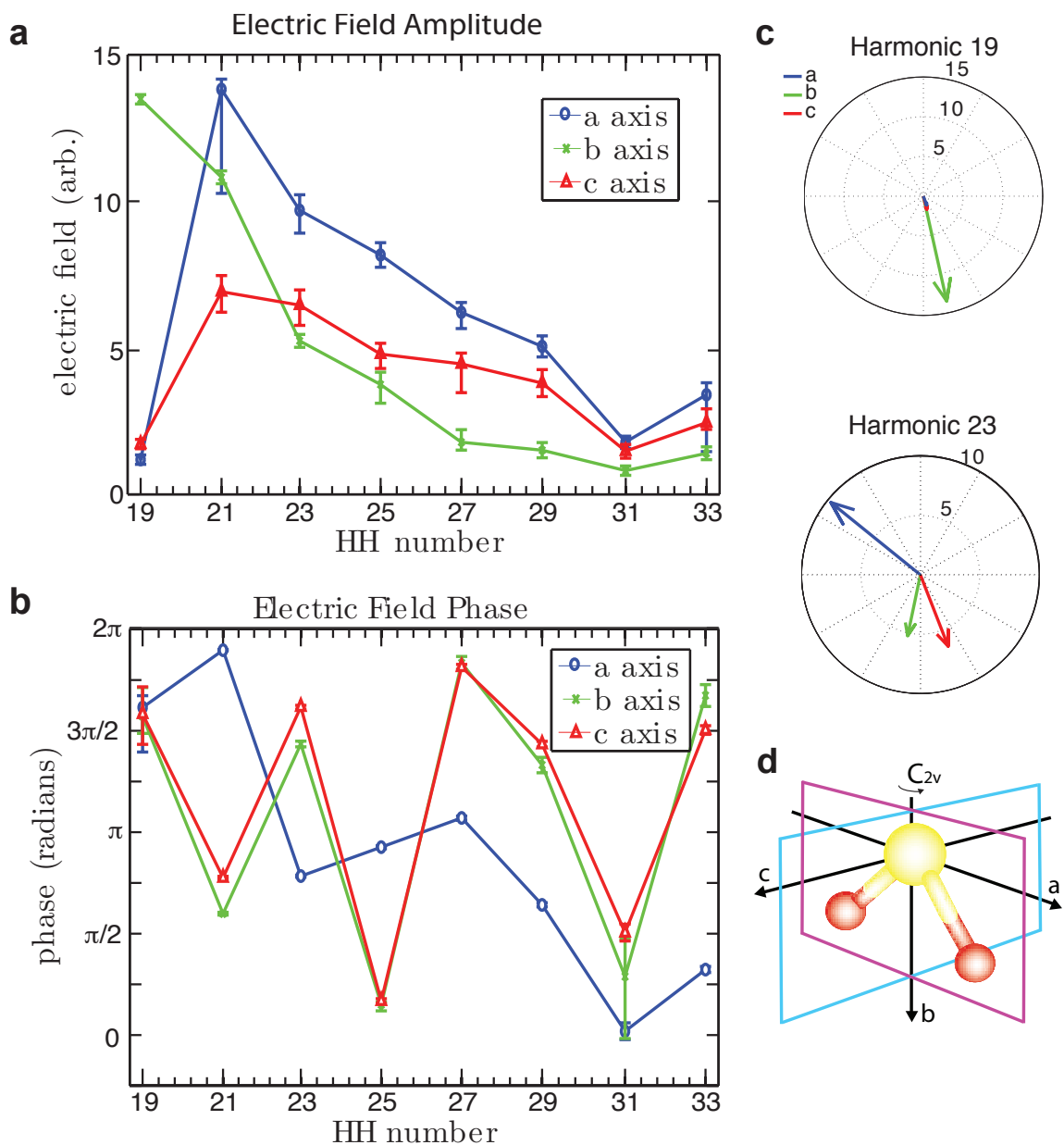

FIGURE 4: Theoretical and experimental contributions to harmonic emission from all three molecular axes across the first few orbitals of $SO_2$ for harmonics 19 and 25. First row: the experiment shows a dominance of the b axis for harmonic 19, but shows the b axis as suppressed by the a axis for harmonic 25. Other rows: calculations show that the HOMO shows a dominance of the b axis for both energies, but the a axis is strong and the b axis suppressed for lower-lying orbitals $5b_2$, $4b_2$ and $7a_1$. Thus, while the 19$^{th}$ harmonic consists mostly of a strong contribution from the HOMO, higher harmonics have stronger contributions from lower-lying orbitals.

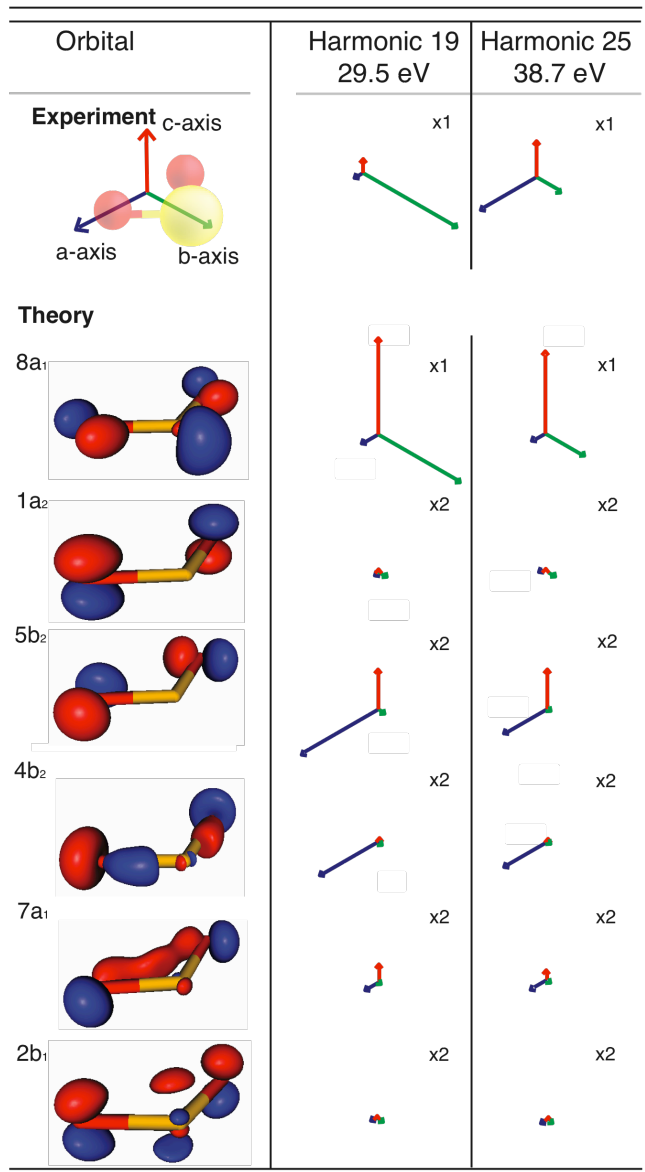

 

\renewcommand{\thefigure}{S\arabic{figure}}



%
\begin{center}
\begingroup\scshape\Large
\bfseries{Supplemental Material}
\endgroup
\end{center}
\vspace{2mm}

%


\section{Orbital contributions to high-harmonic generation in SO$_2$}

In the main text, we discuss the comparison of the theoretical and
experimental contributions to high-harmonic emission using a combination of
the photoionization cross-sections weighted by the field ionization
calculations. However, in several prior papers [35,40-41], it was observed that
the HHG cross spectrum was largely determined solely by the photoionization
cross-section, which models the recombination step of HHG.  As a result, we
examine here a comparison of the theoretical results of photoionization alone
to a combination of photoionization and field ionization calculations. The
high-harmonic power spectrum at the  harmonic energy $\omega$ with the
molecular orientation with respect to the laser-field polarization $\hat{u}$ is modeled using the following equation

\begin{equation} 
S(\omega,\hat{u}) \propto \; \Gamma_{\text{TDSE}}(\vec{u}) \;\times\; \sigma(\omega,\hat{u}),
\end{equation}
where $S$ is the high-harmonic power spectrum, $ \Gamma_{\text{TDSE}}$ is the
ionization yield obtained by the solution of the time-dependent Schr\"odinger
equation (TDSE) describing SO$_2$ exposed to the laser pulse within the one-determinant approximation (see main text) and
$\sigma$ is the total one-photon ionization 
cross-section at a given molecular orientation with respect to the laser polarization.

\begin{figure}[t]
\centering
\includegraphics[height=160mm]{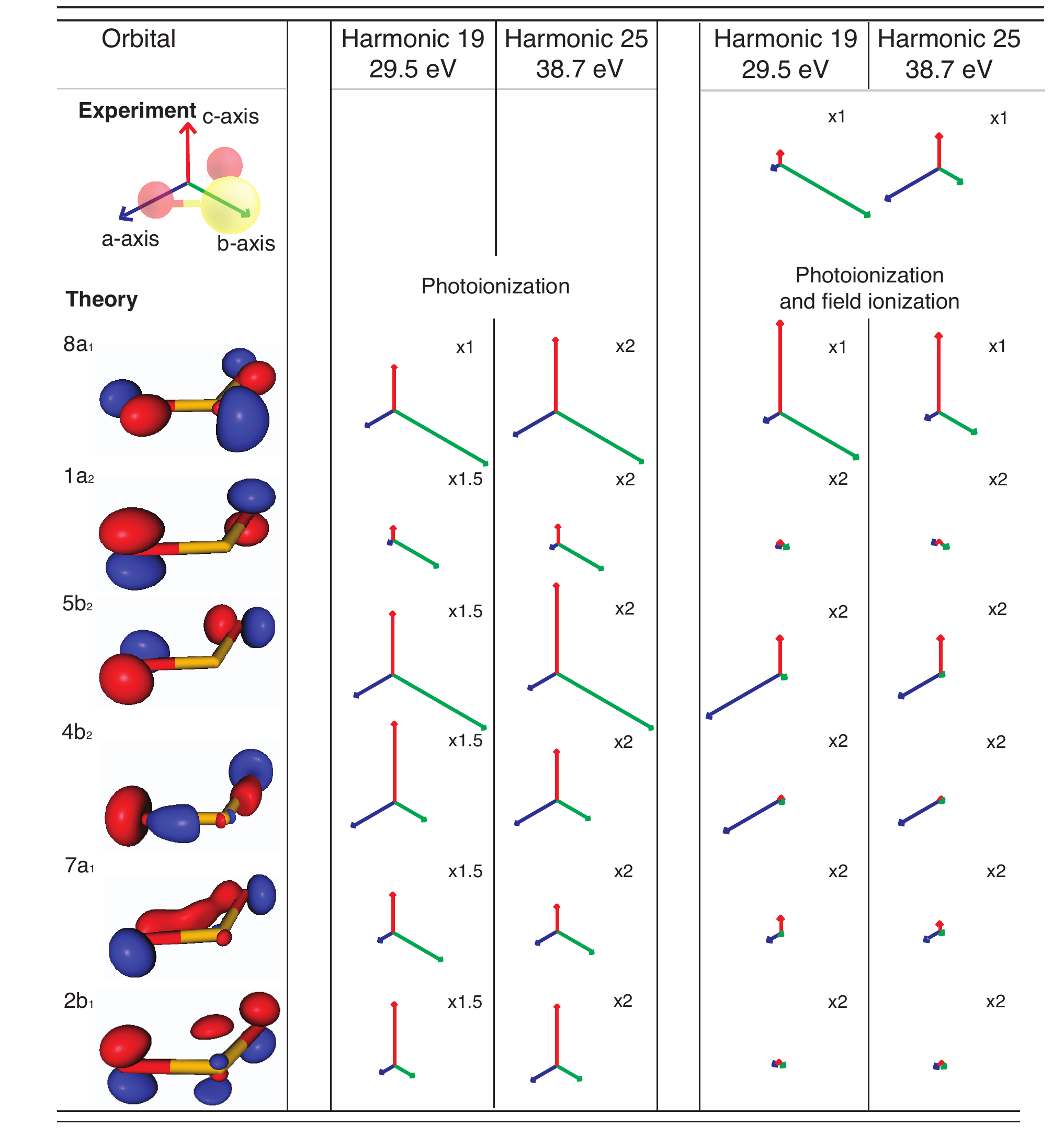}
\caption{Theoretical and experimental contributions to harmonic emission from all three molecular axes across the first few orbitals of SO$_2$ for harmonics 19 and 25. First row: the experiment shows a dominance of the \emph{b} axis for harmonic 19, but shows the \emph{b} axis as suppressed compared to the \emph{a} axis contribution for harmonic 25. Middle columns: photoionization cross-sections, which represent the recombination step of HHG.  The \emph{b} axis provides a strong signal for many of the orbitals, but the \emph{a} axis never shows particular dominance.  Right columns: photoionization cross-sections are combined with field-ionization calculations.  The \emph{a} axis is strong and the \emph{b} axis suppressed for lower-lying orbitals 5b$_2$, 4b$_2$, and 7a$_1$.  We thus find that we require the combined calculations of field ionization and photoionization to explain our data.}
\label{fig:CompTable}
\end{figure}


In Fig.~\ref{fig:CompTable} we show some of the highest-lying occupied orbitals of SO$_2$ (HOMO
through HOMO-5) with the orbital ordering obtained from Ref. [34] and their
contributions to the harmonics 19 and 25. The middle two columns display the relative
contributions from each molecular axis in the experiment (top) and as
calculated for one-photon ionization only (bottom).  The HHG experiment shows the dominance of the \emph{b} axis (green) 
for the 19th harmonic, but a dominant contribution for the \emph{a} axis in
the case of the 25th harmonic (blue).  While the HOMO in the one-photon
ionization calculation shows a strong \emph{b} axis contribution for both the
19th and the 25th harmonics, none of the HOMO or lower-lying orbitals show a
particularly strong contribution from the \emph{a} axis.  The photoionization
calculation on its own is thus insufficient to explain the experiment.

However, looking at the combined photoionization and TDSE strong-field
ionization results, we see that the lower-lying orbitals do show a dominance
of the \emph{a} axis for both harmonic energies.  This points to strong-field
ionization (often in a simplified way referred to as tunneling ionization) as being crucial to fully explain the inversion of signal enhancement present in our experimental data.  We thus find that for some cases, a calculation of the photoionization cross-section alone is insufficient to explain the data and that care should be taken when employing such a calculation on its own.

\begin{figure}[t]
\centering
\includegraphics[height=75mm]{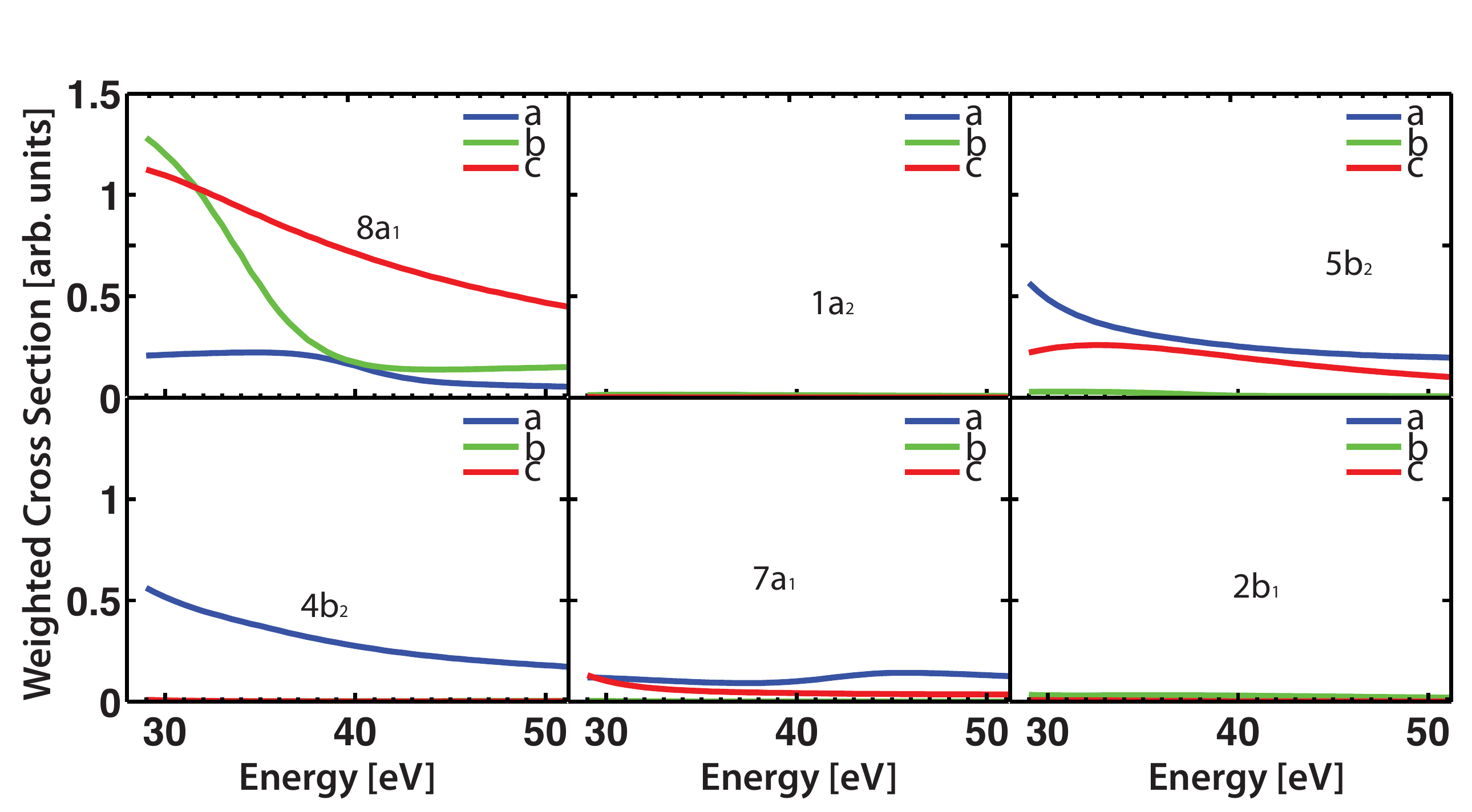}
\caption{One-photon ionization cross-sections weighted by the strong-field ionization yield for the highest-lying occupied orbitals. The lower-lying orbitals 5b$_2$, 4b$_2$, and 7b$_2$ show a dominance of the \emph{a} axis throughout the whole energy range. The lower-lying orbitals contribute more directly to the signal of higher order harmonics.}
\label{fig:AllEnergies}
\end{figure}

\begin{figure}[t]
\centering
\includegraphics[height=100mm]{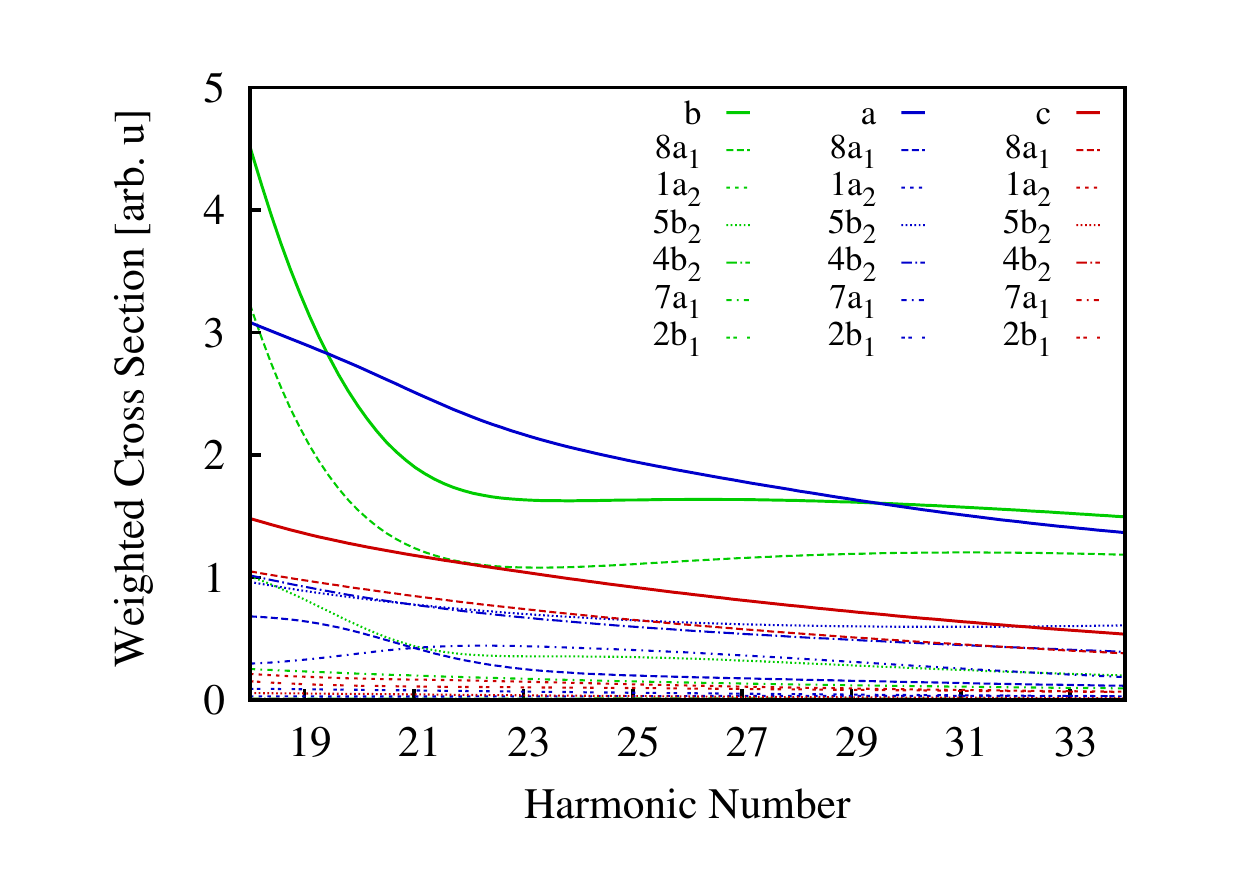}
\caption{Sum of the contribution of the first few orbitals for the main directions of the SO$_2$ one-photon ionization cross-sections weighted by the strong-field ionization yield at an intensity of 2.6 $\times$ 10$^{14}$ W/cm$^2$. The non-solid curves indicate the contribution of each orbital to incoherent sum.}
\label{fig:average_w26}
\end{figure}

For clarity, in the main text we limited our discussion to harmonics 19 and
25.  However, we examined all harmonics to ensure that our analysis was
consistent. In Fig.~\ref{fig:AllEnergies} we show the combined
photoionization and field-ionization results as a function of energy over the
whole energy range of the harmonics 19 to 33. Figure \ref{fig:AllEnergies}
shows that the dominance of the \emph{a} axis that we observe in the
lower-lying orbitals 5b$_2$, 4b$_2$, and 7a$_1$ is consistent throughout the
energy range.  Furthermore, it shows that the HOMO signal switches from being
most pronounced when the laser field is aligned parallel to the \emph{b} axis to
the situation when it is aligned along the \emph{c} axis.  Our analysis of the data shows a switch from the \emph{b} axis at lower harmonics to the \emph{a} axis at higher harmonics as shown in Fig.~3 of the main text. Thus, the explanation that the 19th harmonic receives a dominant signal from the 8a$_1$ orbital while the higher harmonics receive dominant signals from the 5b$_2$, 4b$_2$, and 7a$_1$ is consistent with our analysis of all harmonics.

Fig.~\ref{fig:average_w26} shows the total incoherent sum of the contributions of the different orbitals under study. Here one can see again the dominance of the \emph{b} axis for the harmonic 19 and the dominance of the \emph{a} axis for harmonic 25 due to a crossing in the curve.

\section{Effects of Laser Intensity on Strong-Field Ionization}

Since the experimental laser intensity is a parameter that is not precisely
known, we also consider the effects of laser intensity on the ion yields
obtained when solving the TDSE.  

\begin{figure}[t]
\centering
\includegraphics[width=0.9\textwidth]{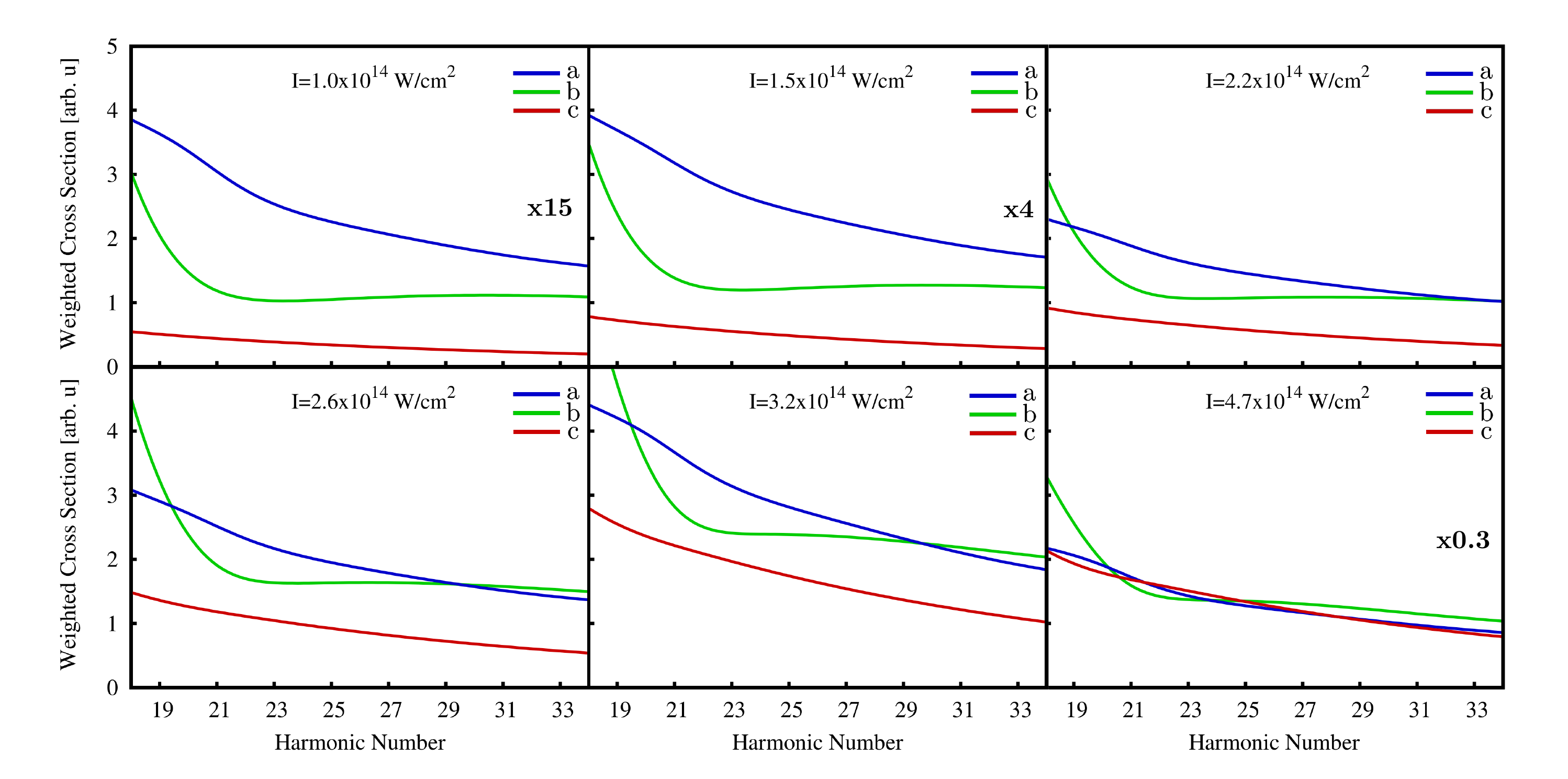}
\caption{SO$_2$ one-photon ionization cross-sections sum weighted by the strong-field  ionization yield at different intensities using the theoretical ionization potentials.}
\label{fig:fusion_teorIP}
\end{figure}

Figure \ref{fig:fusion_teorIP} (similar to Fig.~\ref{fig:average_w26}, but for
better clarity without the orbital contributions) shows the strong-field weighted cross-sections for different peak intensities of the laser. Evidently, the crossing of the blue and green curves (\emph{a} and {b} axes ) moves with intensity, but it exists for quite some intensity range (although it disappears for lower intensities). Interestingly, one sees also that for higher intensities the red curve (\emph{c} axis) increases and shows a crossing with
the green one (\emph{b} axis), as in the experiment. However, we do not see a sharp decrease of the red curves for low HHG numbers, as seen in the experiment for high harmonic 19. Furthermore, one sees also that the blue and the green curves start to approach each other for higher HHG numbers, a similar trend as in the experiment.

\begin{figure}[t]
\centering
\includegraphics[width=0.9\textwidth]{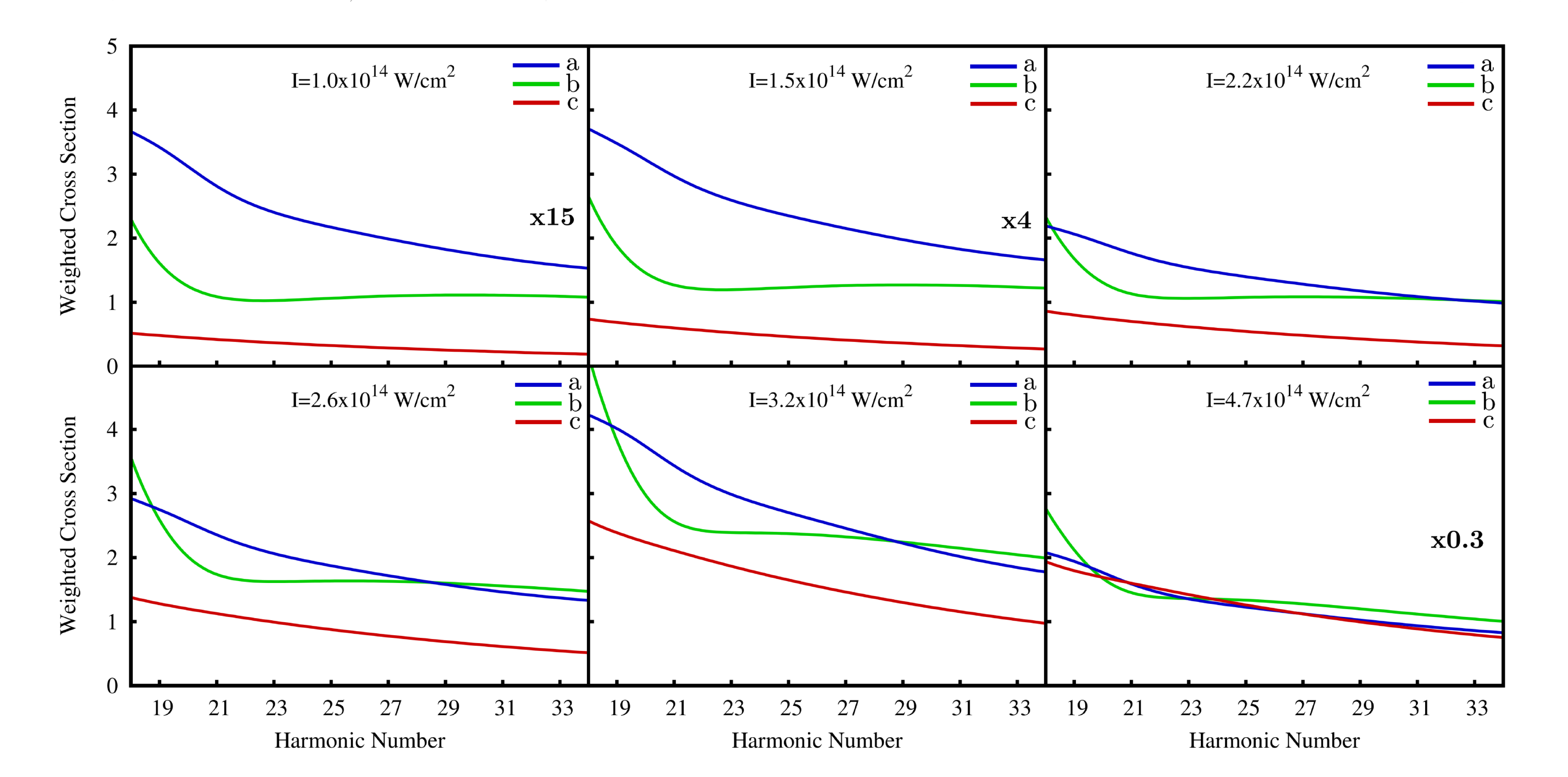}
\caption{Similar to Fig.~S4, but using the experimental ionization potentials}
\label{fig:fusion_expIP}
\end{figure}

Figure \ref{fig:fusion_expIP} shows results for the same parameters as in Fig.~\ref{fig:fusion_teorIP}, but now obtaining the energy axis using experimental ionization
thresholds in the calculation of the one-photon ionization cross-sections. The main finding of the first graph does not change, although the crossing shifts slightly.

We see the crossing due to lower lying orbitals confirmed, and the crossing position lies very close to the experimental one. Furthermore, the crossing and its position is intensity dependent, but it changes rather systematically with intensity and, again, it occurs around the experimental intensity. In fact, the crossing of the green and the red curves seen in the experiment is somehow consistent with the calculation, although the intensities would be higher than the experimentally specified one.

\section{Comparison of the Photoionization Cross Sections}

\begin{figure}[t]
\centering
\includegraphics[height=150mm]{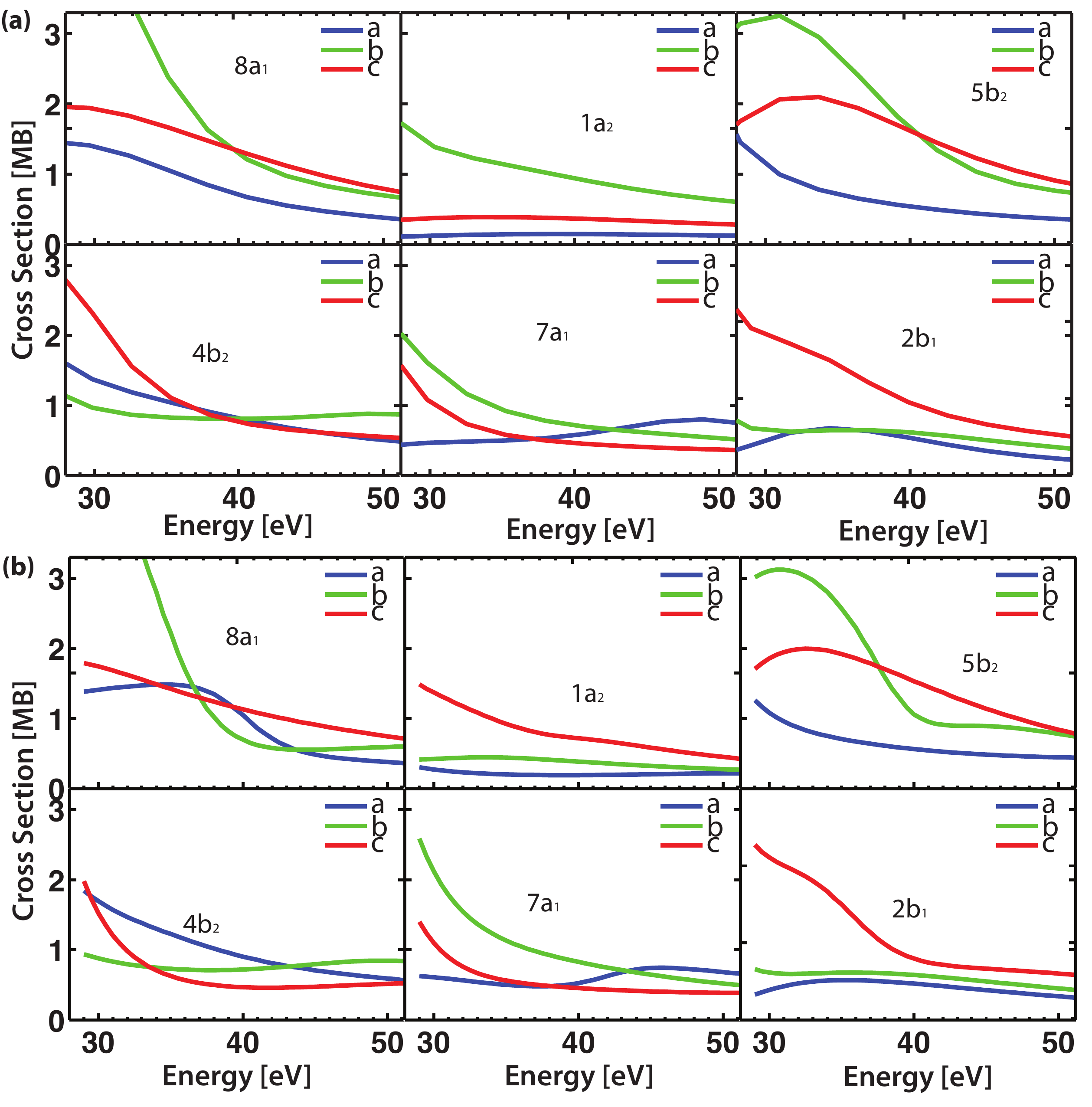}
\caption{A comparison of photoionization cross-section calculations with two different methods. a) Results based on the complex Kohn variational method. These results were used to create the combined results in Fig. 4 of the main text. b) Results obtained using the density-functional theory approach. The minor differences do not influence the overall result when comparing to data or when combined with field-ionization calculations.}
\label{fig:TheoryComp}
\end{figure}

To ensure that the recombination cross-sections were accurately calculated, we
calculated photoionization cross-sections using two different methods, as
mentioned in the main text.  A comparison of the two methods is shown in
Fig.~\ref{fig:TheoryComp}. Fig.~\ref{fig:TheoryComp}a shows results obtained
with the first approach that is based on the complex Kohn variational
method. These results were used in creating the combined probabilities in
Fig.~4 of the main text.  Fig.~\ref{fig:TheoryComp}b shows the results obtained with the second approach.  Although there are some differences between the calculations, they do not influence the overall result when comparing to the data or when combined with strong-field ionization calculations.

\end{document}